\documentclass[onecolumn,12pt]{article}
\usepackage{amsfonts}
\usepackage{bbm}

\usepackage{geometry}
\usepackage{amsmath}
\usepackage{amssymb}
\usepackage{amsthm}
\usepackage{graphicx}
\usepackage{subfigure}
\usepackage{rotating}
\usepackage{lscape}
\usepackage{paralist}
\usepackage{mathrsfs}
\usepackage{lettrine}
\usepackage[mathcal]{eucal}
\usepackage[rm,small,compact]{titlesec}
\usepackage{bibentry}
\usepackage[round,authoryear]{natbib}
\usepackage{abstract}
\usepackage{mathptmx}
\usepackage[scaled=0.92]{helvet}
\usepackage{courier}
\usepackage[T1]{fontenc}
\usepackage{enumitem}
\usepackage{titling}
\usepackage{mathtools}
\usepackage{times}
\usepackage{color}
\usepackage{lineno}
\usepackage{longtable}
\usepackage[varg]{txfonts}
\usepackage[titletoc,title]{appendix}
\usepackage{yfonts}
\usepackage{dsfont}

\theoremstyle{plain}
\theoremstyle{definition}

\newtheorem{lemma}{\textsc{Lemma}}[section]
\newtheorem{theorem}{\textsc{Theorem}}[section]

\newtheorem{proposition}{\textsc{Proposition}}[section]

\newtheorem*{theorem*}{\textsc{Theorem}}
\newtheorem{assumption}{\textsc{Assumption}}[section]

\theoremstyle{definition}
\newtheorem*{remark*}{Remark}

\newtheorem*{assertion*}{\textsc{Assertion}}

\newlist{Axiom}{enumerate}{1}
\setlist[Axiom]{label=Axiom A\arabic*.}

\makeatletter
\renewenvironment{proof}[1][\proofname]{\par
\vspace{-10pt}
    \pushQED{\qed}%
    \normalfont \partopsep=\z@skip \topsep=\z@skip
  \trivlist
  \item[\hskip\labelsep
        \itshape
    #1\@addpunct{.}]\ignorespaces
}{%
    \popQED\endtrivlist\@endpefalse
} \makeatother

 \makeatletter
    \renewcommand{\thefigure}{\ifnum \c@section>\z@ \thesection.\fi \@arabic\c@figure}
    \renewcommand{\thetable}{\ifnum \c@section>\z@ \thesection.\fi \@arabic\c@table}
    \makeatother

\renewcommand{\proofname}{\textsc{Proof}.}

\makeatletter 
\@addtoreset{equation}{section}
\makeatother  

\makeatletter
\newcommand{\rmnum}[1]{\romannumeral #1}
\newcommand{\Rmnum}[1]{\expandafter\@slowromancap\romannumeral #1@}
\makeatother

\makeatletter
\newcommand*\bigcdot{\mathpalette\bigcdot@{.5}}
\newcommand*\bigcdot@[2]{\mathbin{\vcenter{\hbox{\scalebox{#2}{$\m@th#1\bullet$}}}}}
\makeatother

\DeclareSymbolFont{euex}{U}{euex}{m}{n}
\DeclareMathSymbol{\varint}{\mathop}{euex}{"52}

\begin{document}

\setlength{\abovedisplayshortskip}{5pt}
\setlength{\belowdisplayshortskip}{5pt}
\setlength{\abovedisplayskip}{5pt}
\setlength{\belowdisplayskip}{5pt}

\title{\bf \LARGE{Random Discounting and Assessment of Intertemporal Projects: a Non-expected Utility Approach}}

\author{Wei Ma\thanks{Corresponding author: Center for Economic Research, Shandong University, Jinan, 250100, China. Email: wei.ma@sdu.edu.cn.}}

\date{ }
 \maketitle
\begin{onecolabstract}\noindent
This paper assumes each individual in society has a random discount factor and assesses an intertemporal project using rank-dependent expected utility theory. We consider both the ex ante and the ex post approaches. For the former, we show the social planner's discount factor is a convex combination of those of the individuals under the standard Pareto condition. For the latter, we propose a method for determining the social planner's discount factor distribution from the individuals' distributions, which are possibly heterogenous. We demonstrate that relative to expected utility, overweighing of small probabilities can substantially accelerate the decline of the social discount rate.
\end{onecolabstract}

\section{Introduction}\label{section:Introduction}

Collective intertemporal decisions are commonplace in everyday life. Parents need to determine the level of investment on their children's education. A government has to make an intertemporal plan about its oil exploitation. Countries around the world should collaborate to formulate appropriate environmental policies. All these decisions are made by a collective rather than an individual, and are very sensitive to the choice of the social discount rate. For example, on the basis of a near-zero discount rate, \cite{Stern2007} advocates for urgent and immediate actions on climate change, while \cite{Nordhaus2007} argues against this advocation using a market interest rate. Because of this sensitivity, decision makers are usually uncertain about which discount rate to use; rather they entertain a finite number of possibilities, each with a certain probability \citep{Krusell1998, Arrow2013}. This then raises the problem of how to aggregate these discount rate distributions to obtain a social discount rate?\footnote{This problem also has a population interpretation.  The experimental result of \cite{Falk2018} indicates that there is substantial heterogeneity in time preferences both within and across countries. We can represent the heterogeneity within a country by a probability distribution over the discount rates, with the probability of  a discount rate being the fraction of the country's population adopting that discount rate. We are thus led to the problem of how to aggregate these countries' discount rate distributions?}

As an example, consider again the climate change issue. The solution of this issue requires global cooperation. Suppose that, as suggested by \cite{Nordhaus2007}, we use the interest rate as a proxy for the discount rate. It is, however, well documented in the literature that interest rate in almost every nation has experienced substantial fluctuation in the past two decades. Assuming that this historical pattern reflects the likely pattern of interest rate fluctuation in the future, we are faced with the problem of how to aggregate different nations' interest rate fluctuations into a social discount rate in order to assess various climate change policies?

In this paper we adopt a social welfare approach to discounting \citep{Fleurbaey2015}. In risky situations there are two approaches to measure the welfare of a project, one being ex ante and the other ex post. The ex ante approach transforms each random discount rate into a utility function and studies how to aggregate those utility functions under the standard Pareto condition. The ex post approach first aggregates the individuals' preferences over payoff streams with a given discount rate, then forms a probability distribution over all possible discount rates, and finally applies the same decision theory as for the individuals to evaluate a project. In the bulk of the literature, both approaches have been studied within the framework of expected utility (EU) theory. This theory, however,  has been found to be in conflict with a multitude of experimental and empirical facts \citep{Kahneman1979}. For this reason, we invoke one of the most commonly used non-expected utility theory: rank-dependent utility (RDU). This theory has withstood the test of numerous experiments \citep{Quiggin1993}.  

Relative to EU, the most important feature of RDU is the nonlinear probability weighting. To isolate its effect on the social discount rate, we assume all individuals share the same utility function over payoffs, an assumption often made in the aggregation of time preferences \citep{Chambers2018a}, and so  there is no problem of ex post inequality. From this perspective, the present paper is \lq\lq dual\rq\rq to \cite{Fleurbaey2010}, who assumes all individuals share the same probability distribution over states of nature, but have different utility functions over consequences.

We start our analysis with the ex ante approach. Given his distribution for the discount rate, each individual assesses a project according to RDU. In contrast, we do not require the social planner to conform with RDU, but instead assume she is a discounted utility maximizer. We show that under the standard Pareto condition, the discount factor of the social planner is a weighted average of  those of the individuals. Note that in this approach the social planner need not have a probability distribution over the discount rates.

For the ex post approach, since all individuals share the same utility function over payoffs, it is natural to require the social planer to also adopt that function. Therefore, to implement the approach, the key is to ascertain the social planner's probability distribution over discount rates. In the extant literature, the  ex post approach has so far been carried out only when all individuals have a homogenous probabilistic belief \citep{Hammond1981, Broome1990} or the social planner's belief is given a priori  \citep{Fleurbaey2010}. As pointed out by \cite{Mongin2016}, it remains unaddressed how to derive the planner's belief from the individuals' beliefs when they are heterogenous. In the present setting, the individuals' discount rate distributions are heterogenous and we present a method for aggregating these distributions. Specifically, we note that an individual's random discount rate governs his ex post choice behavior, i.e. choice behavior after uncertainty about discount rate is resolved. Such behavior does not depend on how he makes decision under risk. Therefore, we could resort to this  information to develop a method for aggregating the individuals' distributions. 

To obtain such information, we may follow the practice in the experimental elicitation of discount rates by presenting each individual with a set of menus (i.e. a finite set of alternatives) and requesting him to choose one item from the menu after his random discount factor realizes. We ask the individuals to choose from the same menu repeatedly and thereby obtain his choice probabilities for each menu. Note however that the choice probabilities are not completely determined by the distribution of the individual's random discount rate, because it is possible for two items to have the same level of discounted utility for a given discount rate. For this reason, we introduce a tie-breaker: when two items are indifferent, the individual randomly selects another discount factor and picks the one with higher discounted utility. 

To derive the social planner's random discount factor, we impose three conditions: (\rmnum{1}) the social planner's random discount factor depends exclusively on those of the individuals',  (\rmnum{2}) the social planner's tie-breaker is uniquely determined by the individuals' tie-breakers, and  (\rmnum{3}) in any menu, if every individual chooses an item with probability larger (resp. lower) than $1/2$, then so does the social planner. Under these three conditions, we show that if all individuals are free to choose their tie-breakers, the aggregation must be dictatorial; i.e. the social planner's random discount factor is equal to one of the individual's. If, instead, all individuals are demanded to adopt the same tie-breaker, the social planner's random discount factor, under a separability condition, will be a convex combination of those of the individuals'. For expositional simplicity we refer to the latter as the linear aggregation rule. Since the dictatorial structure is unappealing, we assume all individuals adopt the same tie-breaker and the social planner follows the linear aggregation rule. With the distribution for the social planner's random discount rate at hand, the ex post approach can be carried out. 

Now we have obtained a method for implementing both the ex ante and the ex post approaches. It is then natural to ask when the two approaches are consistent, and when they are not, what difference do they make for the social discount rate? We show that the two approaches are consistent if and only if RDU reduces to EU. When they are not consistent, to compare their different effects on the social discount rate, we use the data of \cite{Weitzman2001}. We divide the economists consulted by \citeauthor{Weitzman2001} into two groups. To each group there corresponds a probability distribution over the discount rates. We calculate the social discount rates associated with both the ex ante and the ex post approaches. The numerical example indicates that relative to EU, RDU or nonlinear probability weighting can substantially accelerate the decline of the social discount rate, with the one associated with the ex post approach declining even faster.

This paper contributes to two strands of literature. The first is about the aggregation of time preferences. Historically, it is \cite{Marglin1963} and \cite{Feldstein1964} who first recognize the difficulty in aggregating heterogeneous time preferences. The point is further clarified by \cite{Zuber2011} and \cite{Jackson2015}, who show that a non-dictatorial aggregation of stationary time preferences cannot be simultaneously stationary and Paretian. In a positive note, \cite{Millner2018} show that the aggregated preference can be time-consistent and Paretian, and in the context with multiple generations, \cite{Feng2018} establish that the aggregated preference can be both stationary and inter-generational Pareto. Beyond the utilitarian aggregation method used in the above papers, \cite{Chambers2018a} discuss the non-utilitarian aggregation of exponential discount functions. All these papers study the aggregation of deterministic time preferences. Such preferences, however, are inadequate in capturing individuals' choice behavior, on the one hand because they do not specify how to choose between indifferent alternatives, and on the other hand because a multitude of experimental studies have demonstrated that an individual's intertemporal choice behavior is not deterministic but random \citep{Agranov2017}. For this reason, this paper investigates how to aggregate individuals' random discount factors by exploiting information on their ex post choice behavior.

The second literature is concerned with the aggregation of risk preferences. The ex ante and ex post approaches have been examined extensively in this literature; see, for instance, \citet{Starr1973, Harris1978, Harris1979, Hammond1981, Milne1988}, and references therein. For the ex post approach, it is usually assumed in the extant literature that the individuals and the social planner have a homogenous probability distribution over the states of nature \citep{Broome1990, Fleurbaey2010}. It is not known yet how this approach can be implemented with heterogenous beliefs. The present paper provides such a method.

The paper is structured as follows. In Section~\ref{section: Problem Formulation}, we state the problem and present the ex ante approach. In Section~\ref{section: ex post}, we provide a method for implementing the ex post approach. We compare the two approaches in Section~\ref{section: comparison}. All proofs are collected in the Appendix.

\section{Problem Formulation and the Ex ante Approach}\label{section: Problem Formulation}

Suppose that there are $N$ individuals in society. The objects of choice are called projects, whose payoff streams are represented by $x=(x_0, x_1, \ldots)$ with $x_t\geq 0$, $t=0, 1, \ldots$. Let $\mathcal{X}$ be the set of all projects. We assume each individual (he) is a discounted utility maximizer \`{a} la \cite{Samuelson1937}, but is uncertain about his discount factor and so entertains a finite number of possibilities $\{\beta_1, \ldots, \beta_M\}$ with $M\geq 2$ and $\beta_m\in [0,1]$ for $m=1, \ldots, M$. Let $P^n_m\in [0,1]$ be the probability that individual $n$ adopts the discount factor $\beta_m$, and write $P^n=(P^n_1, \ldots, P^n_M)$.

How should a social planner (she) aggregate these discount factor distributions to obtain a social discount factor?  To answer this question, the ex ante approach proceeds by first constructing a utility function for each individual $n$ based on $P^n$ and then studying the aggregation of these utility functions under the standard Pareto condition. In the literature, the most commonly used utility function is the expected utility function \citep{Hammond1981, Fleurbaey2010}. The EU theory, however, has been found to be in conflict with many experimental and empirical facts \citep{Kahneman1979}. For this reason, we consider in this paper a non-expected utility theory for decision making under risk, i.e. rank-dependent utility (RDU) theory. Specifically, assume without loss of generality that $\beta_1>\cdots>\beta_M$. Define
\begin{equation}\label{equation: RDU}
  U^n(x)=\sum_{m=1}^{M} \left(w_n(\sum_{i=0}^{m}P_i^n)-w_n(\sum_{i=0}^{m-1}P_i^n)\right)\langle \beta_m, x\rangle , x\in \mathcal{X},
\end{equation}
where $w_n: [0,1]\rightarrow [0,1]$ is a probability weighting function (i.e. a continuous and increasing function with $w(0)=0$ and $w(1)=1$), $P_0^n=0$, and $\langle \beta, x\rangle=\sum_{t=0}^{\infty} \beta^t x_t$. Let $U^0_a: \mathcal{X} \rightarrow \mathbb{R}$ be the utility function for the social planner. We assume $U^0_a$ is a discounted utility; that is, there exists a decreasing function $\delta_a: \{0, 1, \ldots\} \rightarrow [0,1]$ with $\delta(0)=1$ such that 
$$U^0_a(x)=\sum_{t=0}^{\infty} \delta_a(t)x_t.$$
The Pareto condition asserts that if for any $x, y\in \mathcal{X}$, $U_n(x) \geq U_n(y)$, $n=1, \ldots, N$, then $U^0_a(x) \geq U^0_a(y)$. 
\begin{proposition}\label{proposition: ex ante}
  Under the  Pareto condition, there exist $\lambda_1, \ldots, \lambda_N$, all being nonnegative and summing up to one, such that 
  \begin{equation}\label{equation: ex ante discount}
  \delta_a(t)=\sum_{n=1}^{N}\lambda_n \sum_{m=1}^{M} \beta_m^t\left(w_n(\sum_{i=0}^{m}P_i^n)-w_n(\sum_{i=0}^{m-1}P_i^n)\right), t=0, 1, 2, \ldots.
  \end{equation}
\end{proposition}

To get an intuition of this proposition, note that \eqref{equation: RDU} can be written as
$$U^n(x)=\sum_{t=0}^{\infty}\left[\sum_{m=1}^{M}\beta_m^t (w_n(\sum_{i=0}^{m}P_i^n)-w_n(\sum_{i=0}^{m-1}P_i^n))\right]  x_t.$$
The term in the square brackets can be understood as the discount factor associated with $U_n$. Given this, Proposition~\ref{proposition: ex ante} says that the social discount factor is a convex combination of the individual discount factors. To see why this is true, note that for a project $x$ and two discount factors $\beta_m, \beta_{m'}$, we have $\langle \beta_m, x\rangle \geq \langle \beta_{m'}, x\rangle \Leftrightarrow \beta_m\geq \beta_{m'},$ and hence $\langle \beta_m, x\rangle \geq \langle \beta_{m'}, x\rangle\Leftrightarrow \langle \beta_m, y\rangle \geq \langle \beta_{m'}, y\rangle$ for any other project $y$. Since RDU satisfies comonotonic independence, it follows that $U_n(\alpha x+(1-\alpha)y)=\alpha U_n(x)+(1-\alpha) U_n(y)$ for $\alpha\in [0,1]$, hence that the range of $(U^0, U^1, \ldots, U^N)$ is convex. The proposition then follows from \cite{DeMeyer1995}. The above discussion provides the essentials for the proof of the proposition, which will therefore be omitted.

\section{The Ex post Approach}\label{section: ex post}

The ex post approach proceeds by first aggregating the individuals' utility functions over the payoff streams, then forming a probability distribution $P^0$ over the $M$ discount factors, and finally applying the same decision theory as for the individuals to evaluate a project. Since we assume all individuals share the same utility function on $\mathcal{X}$, it is natural to require the social planner to also adopt that function. Therefore, to implement the ex post approach, it remains to determine $P^0$. 

Of course, $P^0$ should depend on $(P^1, \ldots, P^N)$. We assume further that $P^0$ depends only on the latter, and $P^0=P$ when all individuals have a homogenous distribution $P$ over the discount factors. Let
$$\Delta=\left\{(P_1,\ldots, P_M): \sum_{m=1}^{M}P_m=1, P_m\geq 0 \text{ for all }m\right\}.$$
\begin{assumption}\label{assumption: function}
  $F: \Delta^N\rightarrow \Delta$ is a function such that $P^0=F(P^1, \ldots, P^N)$ represents the social planner's uncertainty in discount factor when the individuals' uncertainty is given by $(P^1, \ldots, P^N)$, and $F(P, \ldots, P)=P$.
\end{assumption}

What is the appropriate functional form of $F$? In the ex ante approach, we can see from the preceding section that the social planner need not have a probabilistic belief over the set of discount factors, so we cannot use that approach to determine $P^0$. One way out is to resort to the ex post information on an individual's choice behavior, i.e. his behavior after uncertainty is resolved about the discount factor. Specifically, image that an individual is presented with a set of menus (i.e. a finite set of alternatives) and asked to choose one from the menu. Assume the choice is made after his random discount factor realizes and that he is an exponential discounted utility maximizer. Because of the randomness in his discount factor, his ex post choice behavior will also be random. However, this ex post behavior is not completely determined by the individual's distribution over discount factors, because it is possible for two items to be indifferent. Therefore, a tie-breaker is needed.

To break ties, we assume the individual is a lexicographical discounted utility maximizer: when two items tie, he randomly selects another discount factor and picks the one with higher discounted utility \citep[][Supplement]{Gul2006}. Formally, suppose the individual is endowed with a random discount factor $P\in \Delta$. Let $\Omega$ be the set of nonatomic Borel probability measures on $[0,1]$. We call $P\times \upsilon$, or more simply $\upsilon$, a tie-breaker for $P$. Because of the nonatomicity of $\upsilon$, the probability that any two projects have equal discounted utility is zero. To see how $\upsilon$ breaks the tie for $P$, let us consider an example. Take $P$ to be the Dirac measure at $1/2$. Given two projects $x=(1,0, 0,\ldots)$ and $y=(0,2,0,0,\ldots)$, their discounted utilities are the same when the discount factor $\beta=1/2$. To break the tie, we generate another discount factor according to $\upsilon$. The probability of $x$ and $y$ having distinct levels of discounted utility is then given by
$\upsilon\{\beta\in [0,1]: 1\neq 2\beta\}=1.$

Therefore, the tie-breaker $P\times \upsilon$ completely pins down the individual's ex post choice behavior. Specifically, let $\mathcal{D}$ be the set of all menus. We call a function $\rho: \mathcal{X} \times \mathcal{D}\rightarrow [0,1]$ a random choice rule (RCR), in which the value of $\rho(x, D)$ denotes the probability of choosing project $x$ from the menu $D$ and $\sum_{x\in D}\rho(x, D)=1$. It is a complete description of an individual's choice behavior. To relate $\rho$ to $P\times \upsilon$, let
$$M(D, \beta)=\left\{x\in D: \sum_{t=0}^{\infty} \beta^t x_t\geq \sum_{t=0}^{\infty} \beta^t y_t \text{ for all }y\in D\right\}.$$
That is, $M(D, \beta)$ is the set of projects in $D$ which have maximum discounted utility when the discount factor is given by $\beta$. Define
\begin{equation}\label{equation: lexci cone}
\begin{aligned}
  N_l(D,x)&=\{(\beta, \gamma)\in  [0,1]\times [0,1]: x\in M(M(D,\beta), \gamma)\},\text{ and }\\
  \rho(x, D)&=P\times\upsilon(N_l(D,x)).
  \end{aligned}
\end{equation}
It is obvious that $\rho$ is an RCR. We refer to it as the RCR induced by $(P, \upsilon)$.

Let $\upsilon^n$ be the tie-breaker for $P^n$, $n=1, \ldots, N$. We call $(P^1, \ldots, P^N; \upsilon^1, \ldots, \upsilon^N)$ a situation. We need also to determine the tie-breaker for the social planner. For this we assume it depends only on the individuals' tie-breakers.
\begin{assumption}\label{assumption: tie-breaker}
  $G: \Omega^N\rightarrow \Omega$ is a function such that $\upsilon^0=G(\upsilon^1, \ldots, \upsilon^N)$ represents the social planner's tie-breaker when those of the individuals' are given by $(\upsilon^1, \ldots, \upsilon^N)$.
\end{assumption}

With the above preparation, we propose an analog of the standard Pareto condition. Specifically, let $H: \Delta \times \Omega\rightarrow \Gamma$ be a mapping which send each $(P, \upsilon$) to its induced RCR. For notational convenience, let $\mathbf{P}=(P^1, \ldots, P^N)$ be a generic element of $\Delta^N$ and $\Phi=(\upsilon^1, \ldots, \upsilon^N)$ a generic element of $\Omega^N$. Let $\Pi$ be the set of RCRs associated with all  situations; that is,
$$\Pi=\left\{(\rho_1, \ldots, \rho_N; \rho_0): \rho_n=H(P^n, \upsilon^n), \rho_0=H(F(\mathbf{P}), G(\Phi)), (\mathbf{P}, \Phi)\in \Delta^N \times \Omega^N\right\}.$$
Then we propose the following unanimity condition: For all $(x, D)\in \mathcal{X}\times \mathcal{D}$
\begin{equation}\label{equation: Pareto refined}
\begin{aligned}
\forall (\rho_1, \ldots, \rho_N; \rho_0)\in \Pi,  \rho_n(x, D)\geq 1/2,  n=1, \ldots, N \Rightarrow \rho_0(x, D)\geq 1/2,\\
\forall (\rho_1, \ldots, \rho_N; \rho_0)\in \Pi,  \rho_n(x, D)< 1/2,  n=1, \ldots, N \Rightarrow \rho_0(x, D)< 1/2.
\end{aligned}
\end{equation}
That is, if every individual chooses project $x$ from menu $D$ with probability higher (resp. lower) than $1/2$, then so should the social planner. We call $F$ dictatorial if there exists an individual $n$ such that $F(P^1, \ldots, P^N)=P^n$ on $\Delta^N$.
\begin{theorem}\label{theorem: dictatorial}
  If $F$ and $G$ satisfy Assumptions~\ref{assumption: function}, \ref{assumption: tie-breaker}, and condition~\eqref{equation: Pareto refined}, then $F$ must be dictatorial.
\end{theorem}

The dictatorial structure of $F$ is unappealing. To escape it, note that in Assumption~\ref{assumption: tie-breaker}, the individuals are allowed to adopt different tie-breakers. A positive result would obtain, however, if we restrict the assumption by forcing all individuals and the social planner to adopt the same tie-breaker. Specifically,  let
$$\Pi^*=\left\{(\rho_1, \ldots, \rho_N; \rho_0): \rho_n=H(P^n, \upsilon), \rho_0=H(F(\mathbf{P}), \upsilon), (\mathbf{P}, \upsilon)\in \Delta^N \times \Omega\right\}.$$
We restrict \eqref{equation: Pareto refined} to $\Pi^*$: For all $(x, D)\in \mathcal{X}\times \mathcal{D}$
\begin{equation}\label{equation: Pareto refined restricted}
\begin{aligned}
\forall (\rho_1, \ldots, \rho_N; \rho_0)\in \Pi^*,  \rho_n(x, D)\geq 1/2,  n=1, \ldots, N \Rightarrow \rho_0(x, D)\geq 1/2.\\
\end{aligned}
\end{equation}
We call $F$ linear if there exist $N$ nonnegative numbers, $\lambda_1, \ldots, \lambda_N$, summing up to one, such that $F(\mathbf{P})=\sum_{n=1}^{N}\lambda_n P^n$ on $\Delta^N$. To obtain the linear structure, we further assume $F$ satisfies a separability condition. Denote the $k$th component of $F(\mathbf{P})$ by $F_k(\mathbf{P})$
\begin{assumption}\label{assumption: separability condition}
  $F$ is smooth and $\partial^2F_k/\partial P_i^n\partial P_j^{\ell}=0$, $i, j, k=1, \ldots, M$; $\ell, n=1, \ldots, N$ with $\ell\neq n$.
\end{assumption}

In words, this assumption means that each individual cannot influence how others contribute to the social planner's discount factor distribution. 
\begin{theorem}\label{theorem: linear}
  If $F$ satisfies Assumptions~\ref{assumption: function} and \ref{assumption: separability condition}, and condition~\eqref{equation: Pareto refined restricted},  then $F$ is linear.
\end{theorem}

Theorems~\ref{theorem: dictatorial} and \ref{theorem: linear} offer us different ways for ascertaining the probability distribution $P^0$ for the social planner's discount factor. They indicate that how the individuals break ties makes a difference. This is in stark contrast with the ex ante approach, in which the individuals' tie-breaking behavior plays no role. Given $P^0$, the social planner then invokes RDU to assess a project, i.e.
\begin{equation}\label{equation: expost}
U_p^0(x)=\sum_{m=1}^{M} \left(w_0(\sum_{i=0}^{m}P_i^0)-w_0(\sum_{i=0}^{m-1}P_i^0)\right)\langle \beta_m, x\rangle , x\in \mathcal{X},
\end{equation}
where $w_0$ is the probability weighting function for the social planner. It can be seen that $U_p^0(x)$ and $U_a^0(x)$ do not coincide in general.

\section{Comparison of the Ex ante and the Ex post Approaches}\label{section: comparison}
In this section, we examine when the ex ante and the ex post approaches are consistent and the difference between the social discount rates associated with them when they are not consistent.

Let us start with the ex ante-ex post consistency. The ex ante and the ex post approaches are said to be consistent if $U_p^0(x)=U_a^0(x)$ for all $x\in \mathcal{X}$ and all $(P^1, \ldots, P^N)\in \Delta^N$. For the ex post approach, as the dictatorial structure in Theorem~\ref{theorem: dictatorial} is undesirable, we shall in what follows invoke Theorem~\ref{theorem: linear} and assume $P^0=\sum_{n=1}^{N} \lambda_n P^n$, $\lambda_n \geq0$ for all $n$ and $\sum_{n=1}^{N}\lambda_n=1$. For the ex ante approach, we assume $\delta_a(t)$ is given by \eqref{equation: ex ante discount} with the same vector $(\lambda_1, \ldots, \lambda_N)$. 

\begin{proposition}\label{proposition: consistency}
 Assume that $w_n$ are smooth for all $n=0, 1, \ldots, N$. The  ex ante and the ex post approaches are consistent if and only if $w_n(p)=p$ for all $p\in [0,1]$ and all $n$.
\end{proposition}

From the proposition we can see that the ex ante-ex post consistency arises only when RDU reduces to EU. This is in distinction with the results of \cite{Blackorby2004} who establish that the ex ante-ex post consistency fails when $P^1, \ldots, P^N$ are not all the same. The reason is that we assume in the present paper all individuals share the same utility function over payoff streams while \citeauthor{Blackorby2004} allow them to have different utility functions.

Proposition~\ref{proposition: consistency} suggests that we have to make a choice between the ex ante and the ex post approaches if we want to keep nonlinear probability weighting. To facilitate the choice let us examine how differently they affect the social discount rate through a numerical example. We utilize the data of \cite{Weitzman2001}. He consulted 2160 economists, who, by rounding off, suggest $27$ different discount rates in total with three of them being negative and one being zero \citep[see][Table~1]{Weitzman2001}. To avoid technical problems, we discard those three negative rates, and divide the remaining $24$ rates into two groups of equal cardinality, one consisting of those less than or equal to $11\%$ and the other being composed of the discount rates larger than $11\%$. The first group is intended to include the economists who endorse a lower discount rate and the second group a higher discount rate. Within each group, suppose the probability of taking each discount rate is equal to the fraction of the economists in that group who suggest that rate.

\begin{figure}[ht] 
    \centering
    \includegraphics[scale=0.5, bb=11 8 411 335]{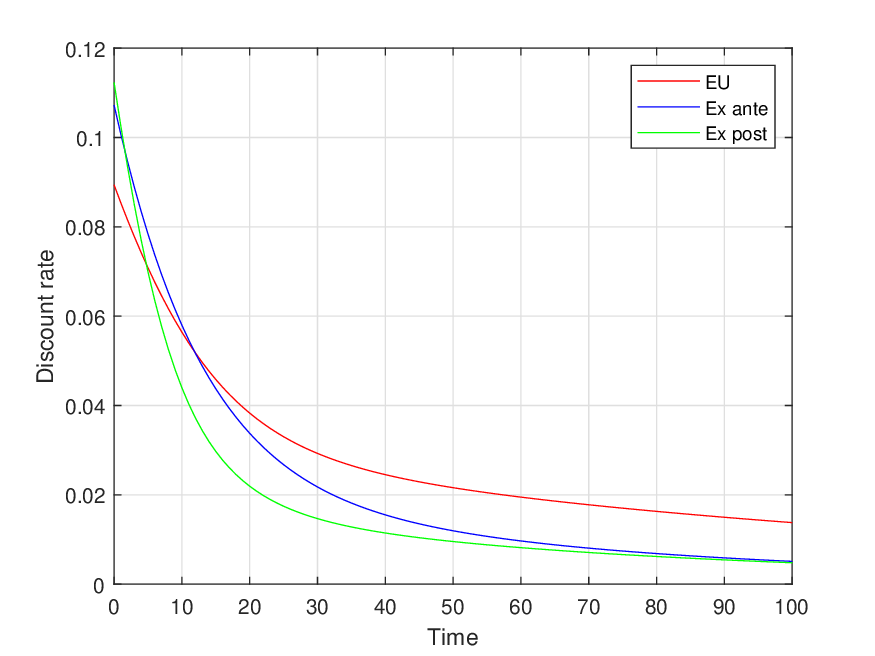} 
   \caption{\footnotesize{Comparison of the discount rates associated with the ex ante and the ex post approaches.}}
   \label{Fig. discount rate}
\end{figure}

We assume that all individuals and the social planner employ the same probability weighting function given by \cite{Gonzalez1999}:
\begin{equation}\label{equation: weighting function}
w(p)=\frac{\theta p^{\gamma}}{\theta p^{\gamma}+(1-p)^{\gamma}}, p\in [0,1],
\end{equation}
where $\theta=0.77$ and $\gamma=0.44$. Take $\lambda_1=\lambda_2=1/2$. Let $\eta_p(t)$ denote the discount rate associated with \eqref{equation: ex ante discount} and $\eta_p(t)$ with \eqref{equation: expost}. Also, let $\eta_e(t)$ denote the discount rate when  the ex ante-ex post consistency holds. With the above specifications, a diagram of the three discount rates against the time period $t$ is presented in Fig.~\ref{Fig. discount rate}. From it we can see that all the three discount rates decline over time, but their declining speed is different: $\eta_e(t)$ declines slowest, $\eta_p(t)$ fastest, and $\eta_a(t)$ in between. Although $\eta_p(t)$ is initially higher than $\eta_a(t)$,  they almost converge when $t=100$, at which time $\eta_e(t)$ is still much higher. This example indicates that nonlinear probability weighting can substantially accelerate the decline of the social discount rate, and that the difference between the discount rates associated with the ex ante and the ex post approaches diminishes fairly fast. As a result, the choice between the two approach will not be an issue of considerable concern when assessing relatively long-term projects. 

Let us make an intuitive and informal discussion about the effect of nonlinear probability weighting function \eqref{equation: weighting function} on the social discount rate. Note that the discount rates associated with expected utility, the ex ante approach, and the ex post approach are given respectively by
$$r_k(t)=\frac{\sum_{m=1}^{M} \alpha_m^k\beta_m^t}{\sum_{m=1}^{M}\alpha_m^k\beta_m^{t+1}}-1, k=e, a, p$$
for some $\alpha_m^k\in [0,1]$, $m=1, \ldots, M$. The values of the numerator and the denominator of the fraction on the right hand side are determined by $\alpha_1^k\beta_1^t$ as $\beta_1$ is the largest discount factor \citep{Weitzman1998}. It is not hard to check that $\partial r_k(t)/\partial \alpha_1^k<0$ and $\alpha_1^e=(P_1^1+P_1^2)/2$, $\alpha_1^a=(w(P_1^1)+w(P_1^2))/2$, $\alpha_1^p=w((P_1^1+P_1^2)/2)$. Since $w(p)$ is concave for small $p$, we have $\alpha_1^p>\alpha_1^a>\alpha_1^e$, hence $r_p(t)<r_a(t)<r_e(t)$. Intuitively, the latter inequalities are due to the social planner's overweighing of small probabilities.

\appendix
\begin{appendices}
\numberwithin{equation}{section}
\section{Proofs}\label{section: Proof}
\subsection{Proof of Theorem~\ref{theorem: dictatorial}}

The proof of the theorem is made by analogy with \cite{Hylland1979} who examine the aggregation of subjective expected utility models. By analogizing each $\beta_m$ to a state of nature, $P^n$ then represents individual $n$'s belief about the occurrence of the states and $\upsilon^n$ his utility function. We first prove an analogy of Lemma~1 of \cite{Hylland1979}, which together with their Lemma~2 gives rise to Theorem~\ref{theorem: dictatorial}.

\begin{lemma}\label{lemma: probability aggregation}
For any two probability profiles in $\Delta^N$, $(P^1, \ldots, P^N)$ and $(Q^1, \ldots, Q^N)$, if $P^n\neq Q^n$ for all $n=1, 2, \ldots, N$, then $F(P^1, \ldots, P^N) \neq F(Q^1, \ldots, Q^N)$.
\end{lemma}
\begin{proof}
We prove by contradiction. Suppose that $F(P^1, \ldots, P^N) = F(Q^1, \ldots, Q^N)$ for two probability profiles $(P^1, \ldots, P^N)$ and $(Q^1, \ldots, Q^N)$ with $P^n\neq Q^n$ for all $n$. By the separating hyperplane theorem \citep[see, e,g., ][Theorem~7.30, p. 276]{Aliprantis2006a}, there exists a vector $a^n=(a^n_1, \ldots, a^n_M)\in \mathbb{R}^M$ such that $P^n\cdot a^n<0$ and $Q^n\cdot a^n>0$. Take $\tau$ to be a sufficiently large positive scalar such that $a^n_m\in (-\tau, \tau)$ for every $m=1,\ldots, M$ and 
\begin{equation}\label{equation: choice of c}
  \tau>\max\left\{(M-2)a_i^n-\sum_{j\neq i}a_j^n, \sum_{i=1}^{M}a_i^n/(M-1), 1\right\}.
\end{equation}
Let $b^n=(a^n+\tau)/2\tau$, so that $b^n_m\in (0,1)$ for all $m$. Then we have
\begin{equation}\label{equation: choice probability reverse}
P^n\cdot b^n<1/2\text{ and }Q^n\cdot b^n>1/2.
\end{equation}

Now we seek to construct $\upsilon^n\in \Omega$ and a menu $D$ with $x\in D$ such that
\begin{equation}\label{equation: choice probability representation}
\rho_n(x, D)=P^n\cdot b^n \text{ and }\eta_n(x, D)=Q^n\cdot b^n,
\end{equation}
where $\rho_n=H(P^n, \upsilon^n)$ and $\eta_n=H(Q^n, \upsilon^n)$. Without loss of generality assume $M\geq 3$ and $1>\beta_1 >\beta_2>\cdots>\beta_M>0$. For otherwise if $M=2$, we can take $\beta_3\notin \{\beta_1, \beta_2\}$ and consider the enlarged set  $\{\beta_1, \beta_2, \beta_3\}$. Let $\beta_0=1$ and $\beta_{M+1}=0$ and take $\gamma_i\in (\beta_{i}, \beta_{i-1})$, $i=1, 2, \ldots, M+1$. Construct the projects
\begin{align*}
  x= & (0, 2, 0, 1, 0, 0,\ldots) \\
  y^m =& (\gamma_{M+1}\beta_m\gamma_m, 2+\gamma_{M+1}(\beta_m+\gamma_m)-\beta_m\gamma_m, \gamma_{M+1}+\beta_m+\gamma_m,  0, 0,\ldots), m=1, \ldots, M.
\end{align*}
Let $D=\{x, y^1, \ldots, y^M\}$. This finishes the construction of project $x$ and menu $D$.

We proceed to construct the tie-breaker $\upsilon^n$. Let
$$c_m^n=\frac{\sum_{j\neq m}b_j^n-(M-2)b_m^n}{M-1}, m=1, \ldots, M.$$
By \eqref{equation: choice of c}, it is not hard to verify that $c_m^n>0$ and $\sum_{m=1}^{M}c_m^n<1$. Then there exists a $\upsilon^n\in \Omega$ such that $\upsilon^n([\beta_m, \gamma_m])=c_m^n$, $m=1, \ldots, M$ and $\upsilon^n([\beta_{M+1}, \gamma_{M+1}])=1-\sum_{i=1}^{M}c_i^n$. We now claim that \eqref{equation: choice probability representation} holds for $\upsilon^n$ and $D$. To see this, take $\rho_n$ for example. Let $\mathcal{P}_x(D)$ denote the collection of subsets of $D$ which contain $x$ and
\begin{equation*}\label{equation: normal cone}
N(D,x)=\left\{\beta\in [0,1]: \sum_{t=0}^{\infty} \beta^t x_t \geq \sum_{t=0}^{\infty} \beta^t y_t \text{ for all }y\in D\right\}.
\end{equation*}
Then
$$N_l(D,x)=\bigcup_{B\in \mathcal{P}_x(D)} \left(\left(  \bigcap_{y\in B} N(D,y) \cap  \bigcap_{y\in D\backslash B} N^c(D,y)   \right) \times N(B,x)\right),$$
where $ N^c(D,y)$ denotes the complement of $N(D,y)$ \citep[see][Supplement, equation~(S1)]{Gul2006}. For each $m=1, \ldots, M$, define
$$\Gamma_m(D,x)=
\begin{cases}
B, &\text{if } \beta_m\in  N(D,y) \text{ for all }y\in B \text{ and } \beta_m\notin  N(D,z) \text{ for all }z\notin B, \\
\varnothing, & \text{otherwise}.
\end{cases}
$$
From the construction of $D$, we have
$$\sum_{t=0}^{\infty} \beta^t x_t - \sum_{t=0}^{\infty} \beta^t y_t^m=(\beta-\gamma_{M+1})(\beta-\beta_m)(\beta-\gamma_m).$$
It follows that $\Gamma_m(D,x)=\{x, y^m\}$ and $N(\Gamma_m(D,x), x)=[0,1]\backslash ([\beta_{M+1}, \gamma_{M+1}]\cup [\beta_m, \gamma_m])$, so that $\upsilon^n (N(\Gamma_m(D,x), x))=\sum_{i\in \{1, \ldots, M\}\backslash \{m\}} c_i^n=b_m^n$. Therefore,
\begin{align*}
\rho_n(x, D)=P^n\times \upsilon^n (N_l(D,x))=\sum_{m=1}^{M} P_m^n \upsilon^n (N(\Gamma_m(D,x), x))=P^n\cdot b^n.
\end{align*}
Let $\rho_0=H(F(P^1, \ldots, P^N), G(\upsilon^1, \ldots, \upsilon^N))$ and $\eta_0=H(F(Q^1, \ldots, Q^N), G(\upsilon^1, \ldots, \upsilon^N))$. Since $F(P^1, \ldots, P^N) = F(Q^1, \ldots, Q^N)$, it follows that $\rho_0=\eta_0$.
By \eqref{equation: choice probability reverse}, we have $\rho_n(x, D)<1/2$ and $\eta_n(x, D)>1/2$ for all $n=1, \ldots, N$, hence, by condition~\eqref{equation: Pareto refined}, $\rho_0(x, D)<1/2$ and $\eta_0(x, D)>1/2$, a contradiction. This proves the lemma.
\end{proof}

Hence, $F: \Delta^N\rightarrow \Delta$ satisfies $F(P, \ldots, P)=P$ and $F(P^1, \ldots, P^N) \neq F(Q^1, \ldots, Q^N)$ for any two probability profiles $(P^1, \ldots, P^N)$ and $(Q^1, \ldots, Q^N)$ with $P^n\neq Q^n$ for all $n=1, 2, \ldots, N$. By Lemma~2 of \cite{Hylland1979}, there exists an $n\in \{1, \ldots, N\}$ such that $F(P^1, \ldots, P^N)=P^n$ for all $(P^1, \ldots, P^N)\in \Delta^N$. This completes the proof.

\subsection{Proof of Theorem~\ref{theorem: linear}}

We begin by showing that for each $\mathbf{P}=(P^1, \ldots, P^N)\in \Delta^N$, there exist $N$ nonnegative numbers, $\lambda_1(\mathbf{P}), \ldots, \lambda_N(\mathbf{P})$, summing up to one, such that $F(\mathbf{P})=\sum_{n=1}^{N}\lambda_n(\mathbf{P}) P^n$ on $\Delta^N$. For this, fix $\mathbf{P}=(P^1, \ldots, P^N)$ and let $P^0=F(\mathbf{P})$. Let
$$\mathcal{C}=\left\{\sum_{n=1}^{N} \lambda_n P^n: \sum_{n=1}^{N}\lambda_n=1, \lambda_n\geq 0,  n=1, \ldots, N\right\}.$$
Suppose by way of contradiction that $P^0\notin \mathcal{C}$. Then there exists a vector $w\in \mathbb{R}^M$ such that
$$P^0\cdot w<0\text{ and }P^n\cdot w>0, n=1, \ldots, N.$$
\citep[See, e,g., ][Theorem~7.31, p. 277.]{Aliprantis2006a} By the argument of Lemma~\ref{lemma: probability aggregation}, we can find a tie-breaker $\upsilon\in \Omega$, a menu $D$, and a project $x\in D$ such that
$$\rho_0(x, D)<1/2\text{ and }\rho_n(x, D)>1/2, n=1, \ldots, N,$$
where $\rho_n$ is the RCR associated with $(P^n, \upsilon)$, $n=0, 1, \ldots, N$. This, however, contradicts with condition~\eqref{equation: Pareto refined restricted}.

We proceed to show that $\lambda_n(\mathbf{P})$ is a constant on $\Delta^N$. By Assumption~\ref{assumption: separability condition}, $F$ must be separable in $P^1, \ldots, P^N$ \citep{Gorman1968}; that is, $F$ takes the form of 
$F(P^1, \ldots, P^N)=\sum_{n=1}^{N}f_n(P^n).$ This implies $\lambda_n(\mathbf{P})=\lambda_n(P^n)$. Since $\sum_{n=1}^{N} \lambda_n(P^n)=1$, we have
$$F(P^1, \ldots, P^N)=\sum_{n=1}^{N-1}\lambda_n(P^n) P^n+(1-\lambda_1(P^1)-\cdots-\lambda_{N-1}(P^{N-1}) )P^N.$$
Direct calculation shows that for all $m=1, \ldots, M$ and $n=1, \ldots, N$,
$$\frac{\partial^2 F_k}{\partial P_m^n \partial P_k^N}=-\frac{\partial \lambda_n}{\partial P_m^n}=0,$$
hence $\lambda_n(P^n)$ must be a constant. This completes the proof.

\subsection{Proof of Proposition~\ref{proposition: consistency}}
For notational convenience, let
\begin{align*}
  a_m&=\sum_{n=1}^{N}\lambda_n \left(w_n(\sum_{i=0}^{m}P_i^n)-w_n(\sum_{i=0}^{m-1}P_i^n)\right),\\
  b_m= &w_0\left(\sum_{n=1}^{N}\lambda_n\sum_{i=0}^{m}P_i^n\right)-w_0\left(\sum_{n=1}^{N}\lambda_n\sum_{i=0}^{m-1}P_i^n\right).
\end{align*}
It follows that
 $$\delta_a(t)=\sum_{m=1}^{M}a_m \beta_m^t,\; \delta_p(t)=\sum_{m=1}^{M}b_m \beta_m^t,$$
where $\delta_p(t)$ is the discount factor associated with $U_p^0$. Since $U_a^0(x)=\sum_{t=0}^{\infty} \delta_a(t)x_t$ and $U_p^0(x)=\sum_{t=0}^{\infty} \delta_p(t)x_t$, then  $U_a^0(x)= U_p^0(x)$ for all $x\in \mathcal{X}$ implies 
$\delta_a(t)=\delta_p(t)$ for all $t=0, 1, \ldots$. Writing the first $M$ equalities in matrix form, we get
$$
\begin{bmatrix}
  1 & 1 & \cdots & 1 \\
  \beta_1 & \beta_2 & \cdots & \beta_M \\
  \vdots & \vdots &\vdots & \vdots \\
  \beta_1^{M-1} & \beta_2^{M-1} & \cdots & \beta_M^{M-1}
\end{bmatrix}
\begin{bmatrix}
  a_1-b_1 \\
    a_2-b_2 \\
  \vdots \\
    a_M-b_M
\end{bmatrix}=0.
$$
Since $\beta_i\neq \beta_j$ for $i \neq j$, it follows that $a_m=b_m$ for all $m=1, \ldots, M$. In particular, consider $a_1=b_1$. This means
$$\sum_{n=1}^{N}\lambda_n w_n(P_1^n )=w_0\left(\sum_{i=1}^{N}\lambda_iP_1^i\right).$$
Differentiating both sides with respect to $P_1^n$, we get
\begin{equation}\label{equation: weighting derivative}
w'_n(P_1^n)=w'_0\left(\sum_{i=1}^{N}\lambda_iP_1^i\right).
\end{equation}
Therefore, $w'_0(p)$ is a constant for all $p\in [0,1]$. To see this, take any two values $p$ and $q$ and consider $(P_1^1, P_1^2, P_1^3, \ldots, P_1^N)=(0, p, 0, \ldots, 0)$ and $(P_1^1, P_1^2, P_1^3, \ldots, P_1^N)=(0, q, 0, \ldots, 0)$. Then taking $n=1$ in the left hand side of \eqref{equation: weighting derivative}, we get $w'_0(p)=w'_0(q)=w'_1(0)$. It then follows that $w'_n(p)$ is also a constant for all $p\in [0,1]$ and all $n=1, 2, \ldots, N$. Since $w_n(0)=0$ and $w_n(1)=1$, we have $w_n(p)=p$, $n=0, 1, \ldots, N$.

\end{appendices}
\bibliographystyle{apa}
\bibliography{library}
\end{document}